# Tuning the electronic structure of graphene by ion irradiation

L. Tapasztó[1], G. Dobrik[1], P. Nemes-Incze[1], G. Vertesy[1], Ph. Lambin[2], and L.P. Biró[1]

1. Research Institute for Technical Physics and Materials Science, H-1525 Budapest, Hungary,
2. Facultes Universitaire Notre Dame de la Paix, 61 Rue de Bruxelles, B-5000 Namur, Belgium

**ABSTRACT**

Mechanically exfoliated graphene layers deposited on $SiO_2$ substrate were irradiated with $Ar^+$ ions in order to experimentally study the effect of atomic scale defects and disorder on the low energy electronic structure of graphene. The irradiated samples were investigated by scanning tunneling microscopy and spectroscopy measurements, which reveal that defect sites, besides acting as scattering centers for electrons through local modification of the on-site potential, also induce disorder in the hopping amplitudes. The most important consequence of the induced disorder is the substantial reduction of the Fermi velocity, revealed by bias dependent imaging of electron density oscillations observed near defect sites.

## I.  INTRODUCTION

Graphene is a remarkable material with excellent electronic [1], mechanical [2] and thermal properties [3]. Its unique low energy electronic properties stem from the massless and chiral behavior of electrons never manifested before in condensed matter physics. However, similarly to conventional nanostructures, graphene is not immune to disorder and its electronic properties are expected to be strongly influenced by the presence of defects. Defects occur predominantly during the production process [4], while an unavoidable source of additional disorder is the interaction with the substrate and environment. However, defects can also be introduced intentionally, i.e. by ion bombardment, in order to engineer the properties of graphene. It was shown theoretically that vacancies can induce two-dimensional magnetic order in graphene [5], as confirmed by experimental observation of proton bombarded graphite [6]. Moreover, several theoretical works pointed out that defects in graphene brings substantial changes in the electronic states near the Fermi level, which states are of acute importance since many of the unique properties of graphene originate from the topology of its electronic bands in the vicinity of the Dirac point [7,8,9,10]. Consequently, it is of particular importance to understand well the physics of defects and disorder in graphene. Scanning Tunneling Microscopy (STM) measurements have proven useful in studying the effects of defect sites of graphite and carbon nanotubes, through imaging of the local density of states (LDOS) in the presence of defect scattering [11,12]. Native defects of bilayer epitaxial graphene grown on SiC have also been investigated [4].

In this paper we report atomic resolution STM and scanning tunneling spectroscopy (STS) measurements on defect sites created by $Ar^+$ ion bombardment of a single graphene layer. We show that, besides acting as scattering centers for electrons, defects also introduce a perturbation and disorder in the hopping amplitudes of the hexagonal lattice, which in turn can seriously alter the Fermi velocity in graphene.

## II. EXPERIMENTAL

The investigated samples consist of mechanically exfoliated graphene flakes deposited on Si substrate with a 300 nm thick $SiO_2$ capping layer. The single layer samples were first identified under an optical microscope, then their thickness checked by atomic force microscopy (AFM) measurements. In order to make them accessible for STM measurements, the graphene flakes were contacted by indium electrodes using a micro-soldering technique recently developed [13]. The main advantage of this contact preparation method is the lack of chemical processing; consequently it enables us to avoid an additional defect-source.

After the contacts were made, the sample was irradiated with $Ar^+$ ions of 30 keV, with a dose of $5 \times 10^{11}$ ion/$cm^2$. The ion dose was chosen to create individual defects in graphene, but with high enough density in order to allow for an easy finding of defects with STM [14]. STM and STS measurements have been performed under ambient conditions both before and after irradiation. The STM tip was positioned under an optical microscope on the contacted graphene flake in order to avoid the crashing of the tip into the insulating $SiO_2$ substrate. Atomic resolution images were typically achieved in constant current mode with the tunneling parameters set to 1 nA and 0.3 V bias.

**III. RESULTS AND DICUSSION**

The atomic resolution image taken on a defect-free region of the irradiated sample shown in Figure 1b clearly displays the honeycomb lattice, characteristic to a single graphene layer in contrast to few-layer graphite flakes where a triangular lattice is imaged by STM [15]. The modulation of the atomic resolution image (brighter/darker areas) is due to the topology (roughness) of the underlying $SiO_2$ substrate as verified by AFM measurements.

Defect sites on the STM images of the irradiated graphene appear as bright spots corresponding to hillock-like protrusions with 0.2 - 0.5 nm height and ~1nm width (Fig. 2.a). Oscillations in the electronic density distribution with a period much larger than that of the graphene lattice can clearly be identified in their close vicinity. The observed oscillations closely resemble the Friedel oscillations calculated for substitutional impurities in graphene [16, 17]. The presence of substituional impurities in our sample is most probable since after irradiation the graphene sample is subjected to ambient conditions. The unavoidable presence of adsorbed water layer can lead to saturation of dangling bonds through dissociation of water at the vacancies created by the ion bombardment [18].

The origin of the oscillations is the interference of the electrons scattered by the defect site [19]. Similar oscillations have been observed near defects in graphite [20, 21] and carbon nanotubes [11, 14] as well as bilayer graphene [4]. The scattering mechanism at the origin of the observed oscillations is the inter-valley scattering of electrons between two non-equivalent K and K' points of the hexagonal Brillouin-zone. In the case of inter-valley scattering a *$2k_F$* change in the momentum of the scattered electrons occurs, where *$k_F$* is the Fermi wave vector, while the oscillation period of the wave function is given by the Fermi wavelength of electrons in graphene ($\lambda_F$=0.74 nm). Since STM maps the square modulus of electronic wave functions the expected periodicity in STM images is 0.37 nm, in good agreement with previous measurements on graphite and CNTs [4,12,20,21]. However, the wavelength of the oscillations measured on irradiated graphene, shown in Figure 2b is about 0.5 nm. The possibility that the observed period is an artifact can be excluded since the image of the atomic lattice has the expected values and is free of distortions.

The observed wavelength can be influenced by altering the distance from the Dirac point (Fermi-level shift) or the slope of the dispersion relation (Fermi velocity). At nonzero Fermi level shift (or decreased Fermi velocity), scattering processes with wave vector changes smaller than $2k_F$ become possible [12], which correspond to a measured spatial periodicity larger than $\lambda_F$. However, such a significant decrease of the Fermi wavelength as observed experimentally would require the shift of the Fermi level by several electron volts, which is unrealistic and is also disproved by the tunneling spectroscopy measurements presented later. The decreased slope of the linear dispersion relation (see Fig 3.a), in turn implies the lowering of the Fermi-velocity. In this latter case, to observe such large wavelength oscillations, we also have to move the Fermi level some way off the bands crossing point but now a few hundreds of meV is already enough to reproduce the observed wavelength, which is an acceptable assumption also confirmed by STS data.

In order to confirm the decrease of the Fermi velocity, we have performed bias dependent STM measurements of the LDOS oscillations, similar to the case measured for carbon nanotubes [11]. The observed oscillation wavelength was found to be dependent on the applied bias potential. This way we could map the dispersion relation near the Fermi level, since, the bias applied for imaging gives the energy relative to the Fermi level, while the corresponding wave-vector can be obtained by measuring the periodicity of the oscillations. This way a linear dispersion relation was revealed (Figure 2.b), as expected for a single graphene layer. The slope of the linear fit gives a Fermi velocity of 3.2 x $10^5$ m/s, which is only one third of the value of 1 x $10^6$ m/s measured for defect free graphene [22]. Here we note that the oscillation periods observed in the STM measurements (in contrast to STS) correspond to energies integrated over the energy window opened by the applied bias potential. However, this averaging only result in a slight overestimation of the Fermi velocity for negative bias potentials applied to the sample (our case). Since in STM images the larger wavelength oscillations are sampled with a stronger weight than those of shorter wavelength, the effect of imaging an energy window is expected to cause a minor overestimation of the measured Fermi velocity [23].

A recent theoretical work shows that the local decrease of Fermi velocity near a substitutional defect sites can be related to the modification of hopping amplitudes from a carbon site to the impurity site as compared to carbon-carbon hopping [8]. For a vacancy the total suppression of hopping to the empty site occurs, however, when an impurity atom is present (i.e. oxygen or hydrogen) we can still have a small but finite hopping probability. The relation between the hopping integral and Fermi velocity in graphene is given by the simple formula: $\hbar v_F = 3ta/2$ where $v_F$ is the Fermi velocity, $t$ the nearest neighbor hopping integral, while $a$ the nearest carbon-carbon atomic distance. For the measured Fermi-velocity the formula gives a hopping amplitude of $t$= 0.9 eV = $0.33t_0$, where $t_0$ is the unperturbed C-C hopping integral ($t_0$ = 2.7eV), indicating the strong suppression of electron hopping from neighboring carbon sites to the impurity site. Here we note that although the modification of hopping amplitudes is a local effect, large oscillation periods (a decreased Fermi velocity) were observed even at several nanometer distances from the defect sites. A possible explanation might be that the fluctuation of nearest-neighbor hopping amplitudes throughout the whole graphene lattice induces a gauge field which can result in a reduced global Fermi velocity as predicted by several theoretical works [24, 25, 26]. An additional source of disorder amplifying the above mentioned effect could be the increased presence of charged impurities (deep traps) in the $SiO_2$ substrate due to irradiation with charged particles [27]. Angle resolved photoelectron spectroscopy measurements have to be performed on irradiated graphene samples in order to fully clarify the local or global aspect of the Fermi velocity reduction.

Theory predicts a specific signature for the substitutional defects of graphene in the tunneling spectra. Namely, when the hopping to the impurity site is suppressed, additional peaks in the STS spectra appear at energies (both below and above the Fermi level) close to the value of the decreased hopping integral [8]. In order to confirm our interpretation for the decreased Fermi-velocity, and validate the theoretical prediction of Pereira et al. [8] we have performed STS measurements at the defect sites. The measured STS spectra compared with that of defect free graphene is shown in Figure 3b. Additional peaks appear at ±1 eV near the Dirac point as compared to defect-free graphene spectra. Conform to the theory these peaks correspond to a hopping integral value of $t_0 \approx 1eV$. The decreased hopping amplitude value ($t_0$) resulting from STS measurements is in good agreement with the value of 0.9 eV given by the analysis of energy resolved Freidel oscillations measured by STM.

Furthermore, the STS spectra is asymmetric to the zero bias voltage, which reveals that the Fermi level is shifted by an amount of about -400 meV from the bands crossing (Dirac) point. We attribute the Fermi-level shift primarily to the charge transfer between graphene and indium contacts. Indium has a work function of 4.12 eV, while the work function of graphene is about 4.5 eV. The aligning of the Fermi levels by transfer of electrons from In contacts to graphene result in an n-doped graphene with an asymmetric position of the band structure relative to the Dirac point. The magnitude of the Fermi-level shift is in good agreement with the theoretical prediction [28], which gives a shift value of about -450 meV for indium on graphene. As a further check of consistency, we have extrapolated our measured dispersion relation to zero bias (energy), which gives a Fermi level shift of about -440 meV which is also in good agreement with the shift observed directly from STS spectra.

The above results provide strong experimental evidence for the suppression of the hopping amplitudes to substitutional defect sites and validate the theoretical prediction concerning their effect on the Fermi velocity.

## IV. CONCLUSIONS

To summarize, by performing atomic resolution STM and STS measurements on graphene samples irradiated with $Ar^+$ ions we showed that beside the well-known modification of the local on-site potential (electron scattering), substitutional defects in graphene also induces a disorder in the hopping amplitudes, which can substantially alter the Fermi velocity. This way we can tune the Fermi velocity of graphene by intentional introduction of defect sites by ion irradiation, which could open up new perspectives for graphene electronics.

**ACKNOWLEDGMENTS**

Financial support from OTKA-NKTH grant 67793 and OTKA-NKTH grant 67851 is acknowledged.

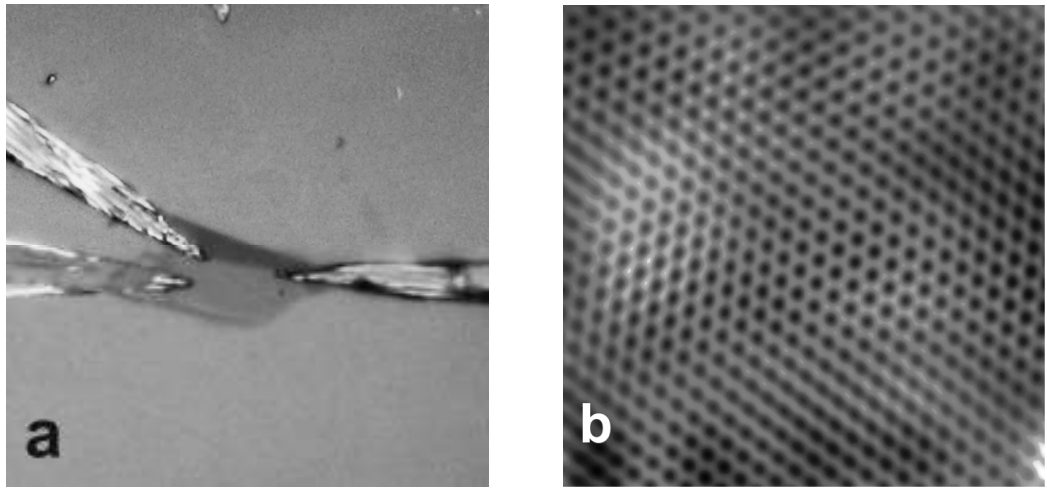

Fig .1. a) Optical microscopy image of a graphene flake on Si/SiO$_2$ substrate contacted by indium electrodes using the micro-soldering method. b) Atomic resolution STM image (6 x 6nm, 1nA, 100mV) of graphene on SiO$_2$, displaying the honeycomb carbon lattice characteristic for single graphene layers.

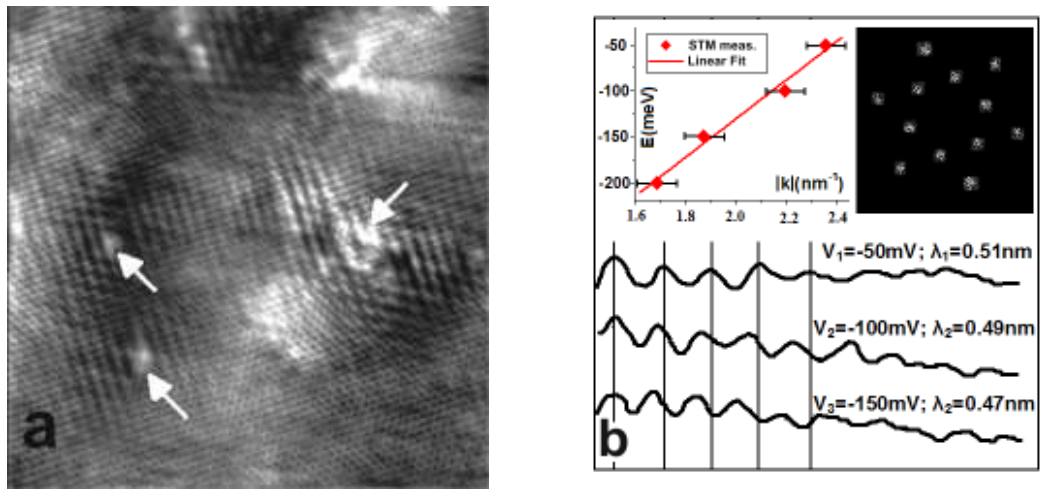

Fig. 2.a) Atomic resolution STM image (20x20 nm, 1nA, 100 mV) of irradiated graphene on SiO$_2$, revealing electron density oscillations near defect sites (small hillock-like protrusions indicated by arrows). b) Line cuts on STM images of oscillations recorded near the same defect

site at different bias voltages. Experimental dispersion relation revealing a strongly suppressed Fermi velocity (upper left inset) and 2D Fourier transformed STM image displaying two hexagons; the larger corresponding to the periodicity of honeycomb structure, while the smaller one to the observed larger periodicity oscillations (upper right inset).

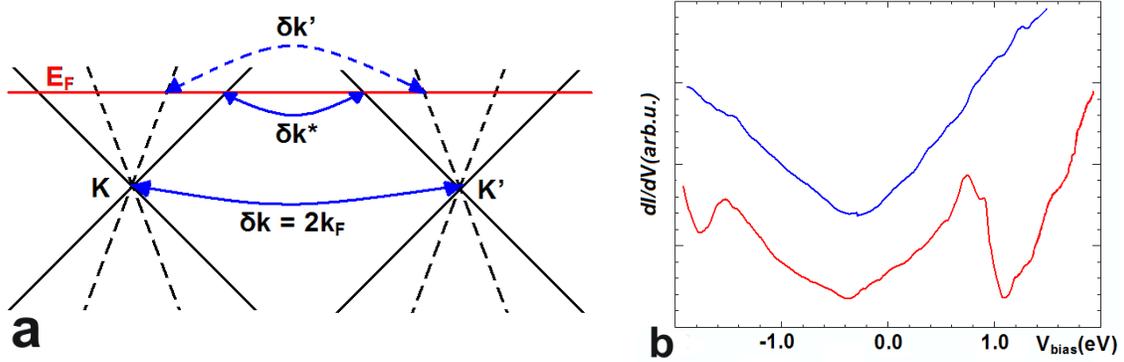

Figure 3. a) Dispersion relation near two nonequivalent Dirac points K and K' of the Brillouin zone for irradiated (full) and defect-free (broken line) graphene samples. The inter-valley scattering processes at the shifted Fermi level (horizontal line) are shown for the case of normal (δk') and reduced (δk*) Fermi velocities. b) Scanning tunneling spectra of graphene taken on the defect free region (upper) and at a defect site of the irradiated graphene (lower).